\documentclass[useAMS,usenatbib]{mn2e}
\usepackage{times}
\usepackage{graphics,epsf}
\usepackage{amsmath}                
\usepackage{amsfonts}               
\usepackage{amssymb}                
\usepackage{epsfig}                 
\usepackage{rotating}
\usepackage{color}
\usepackage{tabularx}

\newcommand{\cm}{{~\rm cm}}
\newcommand{\s}{{~\rm s}}
\newcommand{\km}{{~\rm km}}
\newcommand{\g}{{~\rm g}}
\newcommand{\K}{{~\rm K}}
\newcommand{\erg}{{~\rm erg}}
\newcommand{\yr}{{~\rm yr}}

\newcommand{\pc}{{~\rm pc}}

\newcommand{\AU}{{~\rm AU}}

\newcommand{\days}{{~\rm day}}

\def \apj{ApJ}
\def \aap{A\&A}

\def \mnras{MNRAS}
\def \apjl{ApJ Lett.}
\def \apjs{ApJ Suppl. Ser.}
\def \nat{Nature}
\def \actaa{Acta Astron.}

\def \nar{New A Rev.}
\def \planss{Planet.~Space~Sci.}

\begin{document}

\title{What sodium absorption lines tell us about type Ia supernovae}

\author[Noam Soker] {Noam Soker\\ Department of Physics, Technion -- Israel Institute of Technology, Haifa 32000, Israel;\\soker@physics.technion.ac.il}
\maketitle


\begin{abstract}
We propose that the sodium responsible for the variable Na~I~D absorption lines in some type Ia supernovae (SN Ia) originate {{{{  {mainly} }}}}
from dust residing at $\sim 1 \pc$ from the supernovae.
In this Na-from-dust absorption (NaDA) model the process by which the SN Ia peak luminosity releases sodium from dust at $\sim 1 \pc$ from the
SN is similar to the processes by which solar radiation releases sodium from cometary dust when comets approach a distance of $\la 1 \AU $ from the Sun.
The dust grains are not sublimated but rather stay intact, and release sodium by photon-stimulated desorption (PSD; or photo-sputtering).
{{{{  {Some of the Na might start in the gas phase before the explosion. Weakening in absorption strength is caused by Na-ionizing radiation of the SN. }  }}}}
We apply the NaDA model to SN~2006X and SN~2007le, and find it to comply better with the observed time variability of the Na~I~D absorption lines than the Na recombination model.
The mass in the dusty shell of the NaDA model is much too high to be accounted for in the single-degenerate scenario for SN Ia.
Therefore, the presence of variable Na~I~D lines in some SN Ia further weakens the already very problematic single-degenerate scenario for SN Ia.
\end{abstract}


\section{INTRODUCTION}
\label{sec:introduction}

The variety of scenarios (e.g., \citealt{WangHan2012}) for the formation of type Ia supernovae (SN~Ia) can be classified in several different ways,
one of which is the following.
\newline
(a) \emph{The double degenerate (DD) scenario} (e.g., \citealt{Webbink1984, Iben1984}).
Two white dwarfs(WDs), with a combined super- or sub-Chandrasekhar mass (e.g., \citealt{vanKerkwijk2010, BadenesMaoz2012}),
lose angular momentum and energy through the radiation of gravitational waves \citep{Tutukov1979}
and merge. The time from merger (or onset of mass transfer) to explosion is unknown (e.g., \citealt{vanKerkwijk2010}).
\newline
(b) \emph{The core-degenerate (CD) scenario} \citep{Livio2003, KashiSoker2011, IlkovSoker2013, Sokeretal2013, Sokeretal2014}.
A WD merges with a hot core of a massive asymptotic giant branch (AGB) star.
The explosion can occur shortly after the common envelope phase, hence leading to a
SN Ia inside a planetary nebula shell \citep{TsebrenkoSoker2013, TsebrenkoSoker2014}, or at a very long time delay \citep{IlkovSoker2012}.
There is some overlap between the DD and CD scenarios.
\newline
(c) \emph{The single degenerate (SD) scenario} (e.g., \citealt{Whelan1973, Nomoto1982, Han2004}).
A WD reaches the Chandrasekhar mass limit by accreting a hydrogen-rich mass from a non-degenerate stellar companion, and explodes.
If the accreted mass is helium-rich (e.g., \citealt{Iben1987, Ruiter2011}),
this scenario can be listed under the double-detonation scenario.
\newline
(d) \emph{The `double-detonation' mechanism.} (e.g., \citealt{Woosley1994, Livne1995}),
A sub-Chandrasekhar mass WD accumulates a layer of helium-rich material that detonates and set a second detonation near the center of the CO WD
(e.g., \citealt{Shenetal2013}).
\newline
(e) \emph{The WD-WD collision scenario} (e.g., \citealt{Thompson2011, KatzDong2012, Kushniretal2013}).
A tertiary star brings the two inner WDs to collide at about their mutual free-fall velocity and immediately explode.
Despite some attractive features, this scenario can account for at most few per cent of all SN Ia \citep{Hamersetal2013, Prodanetal2013, Sokeretal2014}.

As there is no consensus on the evolutionary routes of SN Ia (e.g., \citealt{Livio2001, Maoz2010, Howell2011, WangHan2012, Maozetal2014}), and
each scenario suffers from one or more (severe) problems \citep{Sokeretal2014}\footnote{also Soker, N. 2013: http://grb.physics.ncsu.edu/FOE2013/WEB/abstracts.html},
any observation is critical to better constrain at least some of the scenarios.
One such observation is the neutral sodium D absorption lines \citep{Patatetal2007, Sternbergetal2011, Sternbergetal2013, Simonetal2009}.
Some studies (e.g., \citealt{Patatetal2007, Patatetal2011, Sternbergetal2011, Phillipsetal2013, Simonetal2009, Borkowskietal2009, Boothetal2014}) attributed the Na~I~D absorption lines to a wind from
a giant star in the SD scenario.
Attributing any CSM to the SD scenario was criticized by \cite{Sokeretal2013} who showed that at least for SN PTF~11kx the circumstellar material (CSM)
is much too massive to be accounted for in the SD scenario.

In a recent paper \cite{Sternbergetal2013} did not detect time variability in Na~I~D absorption line strength that can be associated with CSM of the 14 SN Ia they have studied.
Motivated by this null detection we reexamine the case of SN~2006X, for which \cite{Patatetal2007} found a strong Na~I~D absorption lines
varying over a time scale of weeks.
In section \ref{sec:SN2006X} we find some problems in the simple model of \cite{Patatetal2007} for SN~2006X.
Instead, we suggest a model where the sodium is released from dust, much as in tails of comets within $\sim 1 \AU$ from the Sun.
In section \ref{sec:SN2007le} we apply the Na-from-dust absorption (NaDA) model to SN~2007le.
We summarize in section \ref{sec:Summary}.

\section{THE CASE OF SN~2006X}
\label{sec:SN2006X}

\cite{Patatetal2007} observed the evolving absorption Na~I~D lines in SN~2006X and built a model where the SN ionizes the neutral sodium, which then recombines.
After recombination absorption by atomic Na increases.
As the absorbtion appears within $\sim 10~$days, the recombination time in their model, the electron density should be $n_e \sim 10^5 \cm^{-3}$.
The hydrogen in the gas, which supplies most of the electrons, must be ionized by the supernova radiation.
The SN ionizing-photon flux around peak luminosity is $S_{\rm UV}=4.4\times10^{44}~$photon$\s^{-1}$, and it lasts for
$\Delta t_{\rm SN} \sim 20~$days.
\cite{Patatetal2007} assumed that the shell thickness is $\Delta r =0.1r$, where $r$ is the shell distance from the explosion, and concluded that the shell
must reside within $r_H \la  10^{16} \cm$.
\cite{Simonetal2009} criticized this model and based on ionization considerations argued that the absorbing gas must reside at $r > 10^{16} \cm$.
\cite{Chugai2008} modelled the absorption by Na and Ca from AGB winds, and found that the expected optical depth in the Na~I~D $5890 \AA$ line is very low, $\tau < 10^{-3}$.
{{{  {We also note that \cite{CrottsYourdon2008} study the light echo from SN~2006X and find little evidence for CSM. They argue that most of the echoing
material resides at $26 \pc$ from the SN.} }}}

Although the model of \cite{Patatetal2007} was already criticized by \cite{Chugai2008} and \cite{Simonetal2009}, we here list additional problems in the model
in order to set the stage for our proposed NaDA model.
\newline
(1) {\it Large velocity spread.} The absorption line spreads over a velocity range of $-50 \km \s^{-1}$ to $+50 \km \s^{-1}$.
It is not clear how a large spread of $\sim 100 \km \s^{-1}$ could have been formed from a wind of a giant star within such a narrow shell
{{{  {at a large distance from the star. Gas parcels moving with velocities differences as large as their outflow speed, will catch each other close to the star.
I attribute the similar velocity spread in RS Ophiuchi \citep{Patatetal2011} to gas motion near the binary system. Indeed, the outflow speed of gas in
RS Oph is $<100 \km \s^{-1}$, such that when observed two years after the nova outburst it was at a distance of $<10^{15} \cm$,
much smaller than the location of the absorbing gas in SN~2006X.
In SN~2006X the Ca lines are not variable, while in RS Ophiuchi they are. This further suggests a different behavior in the two systems.
\cite{Mohamed etal2013} simulate a SN Ia inside an RS Oph type system, and found the Na absorption lines to decay almost completely after one months,
contrary to the long lasting deep absorption lines in SN~2006X. }}}
Several shells can in principle be formed by nova outbursts, but then it is hard to explain their formation at different velocities and their dense spacing within
$\sim 5 \times 10^{16} \cm$.
{{{   {\cite{Boothetal2014} simulated the formation of such shells by a symbiotic system. In their simulation there is a spiral structure in the equatorial plane.
But all segments of the spiral structure have the same velocity, and the density is much lower than that required by the model of \cite{Patatetal2007}
at $r \ga 5 \times 10^{15} \cm$. } }}}
\newline
(2) {\it Ionization time scale.} At day 14 after peak some velocity segments, e.g., $v=20 \km \s^{-1}$ in their figure 1,
reach complete recombination according to their model.
At that time the ionizing radiation is still large, e.g., the luminosity in the U band declined to only $\sim 30 \%$ of peak luminosity \citep{Wangetal2008}.
It is not clear how the sodium at $v\sim 20-50 \km \s^{-1}$ reached
complete recombination, as pointed out also by \cite{Chugai2008} and \cite{Simonetal2009}.
\newline
(3) {\it Interaction with the ejecta.} The ejecta of SN~2006X has a maximum velocity of $v_{\rm exp} \simeq 20,000 \km \s^{-1}$
\citep{Quimbyetal2006}.
If indeed the sodium absorbing gas resides at $r_H < 4 \times 10^{15} \cm$, then the front of the supernova
interacts with it and destroys it at about 23~days from explosion, or in less than 10 days after maximum.
{{{   {\cite{Patatetal2007} consider some of the CSM in their model to collide with the ejecta (also \citealt{Boothetal2014}), as might be the case for the
lower blue-shifted part whose absorption diminished between days +14 and +61 (Fig 1 of {\citealt{Patatetal2007}). However,} }}}
most of the velocity spread in absorption spectrum seems to be unaffected up to day +121 \citep{Patatetal2007}, hence must reside at
$r \ga 2 \times 10^{16} \cm$.
\cite{Patatetal2007} allows for such a shell if the SN hydrogen ionizing flux is much larger, by few hundreds, than what they take in their equation (S2).
However, for $r \ga 2 \times 10^{16} \cm$ the mass in the shell is $\sim 0.001 M_\odot$ (their equation S4) and the mass loss rate from the giant according to their equation (S5)
is $\sim 10^{-5} M_\odot \yr^{-1}$, much higher than in symbiotic stars, and time between nova outbursts in their model is $\sim 100 \yr$.
For such a mass loss rate and a wind velocity of $20 \km \s^{-1}$ the hydrogen number density at $r=3 \times 10^{15} \cm$ is $\sim 10^6 \cm^{-3}$,
and we are back in forming a shell at $\sim 10^{15} \cm$. The SN itself will be enshrouded in dense wind, probably containing dust.

We suggest that the {{{{  {most of the} }}}} sodium originates from photon-stimulated desorption (PSD; or photo-sputtering) of dust grains
residing at $r_d \sim 1 \pc$ from the exploding WD.
The dust grains are not destroyed as their temperature at that distance from the explosion is much below their sublimation temperature.
The substantial extinction of SN~2006X \citep{Quimbyetal2006} might come in part from a dust residing at that distance.
Another SN Ia with variable Na~I~D absorption line is SN~1999cl \citep{Blondinetal2009}.
Both SN~1999cl and SN~2006X are the two most highly reddened objects in the sample of \cite{Blondinetal2009},
who mentioned that the variability is connected to dusty environments.
In our suggested scenario the number of neutral sodium atoms was increasing near peak luminosity, leading to the increased absorption.
In the Na~I~D line velocity range of $20-50 \km \s^{-1}$ the absorption was high, but then decreased (figure 1 in \citealt{Patatetal2007}).

{{{  {Re-adsorption of Na atoms on dust grains seems to be too slow to account for this variation.
The weakening absorption in this velocity range is attributed in the NaDA model to increase in ionization fraction of Na with time after peak luminosity,
as in the calculation of \cite{Borkowskietal2009}. \cite{Borkowskietal2009} study the variation in Na~I absorption due to ionization of a shell at $\sim 1-10 \pc$.
Because of the low density recombination is slow, and their model can account for  weakening absorption, but not to increase absorption.
In the NaDA model the PSD of Na increases absorption and ionization decreases absorption. The competition between these two opposite effects determine the
variation in Na~I absorption strength.
Some, or all, of the Na in this velocity range might be in the gas phase before the explosion.
This can be the case if this velocity-range gas resides closer to the star, but not as close to have been run over by the ejecta. } }}}

We based our scenario on the known behavior of dust in the solar vicinity.
The presence of sodium from sublimated grains is expected at $\sim 10 R_{\sun}$ from the Sun (e.g., \citealt{Deloneetal2008}).
With a peak luminosity of $\sim 10^{10} L_{\sun}$ we expect dust to be sublimated at $r_s \sim 0.02 \pc$ from the SN with partial sublimation further out \citep{Kochanek2011},
up to $\sim 0.05 \pc$ \citep{Simonetal2009}.
We look at dust further out.
Comets are known to release atomic sodium from dust at distances up to $1.4 \AU$ from the Sun, with comet Hale-Bopp detected sodium tail at $0.98 \AU$ from the Sun
as a prominent example \citep{Cremoneseetal1997}.
Scaling to the luminosity of SN~2006X, sodium might be released from dust up to a distance of $\sim {\rm few} \times10^5 \AU \sim 1 \pc$.
\cite{Leblancetal2008} studied the comet McNaught~C/2006~P1 and conclude that the ejection of Na atoms
is most probably related to photo-sputtering (PSD) of the dust grains ejected from the comet.

\cite{Chugai2008} already proposed that the strong Na and weak Ca absorption can be attributed to dense dusty clouds residing beyond the sublimation radius of dust,
where Na~I is present but Ca is depleted onto dust grains.
He attributed the time variation to the presence of several clouds, or partial coverage together with photosphere expansion.
We attribute the time variation to the release of Na, increasing absorption,  and {{{{  {to ionization of Na that reduces absorption.} }}}}

\section{THE CASE OF SN~2007le}
\label{sec:SN2007le}

\cite{Simonetal2009} built the following model for the variable Na~I~D lines in SN~2007le.
They assumed a hydrogen (atoms+protons) density of $n_H=2.4\times 10^7 \cm^{-3}$ and calculated the fraction of atomic Na in a cloud at
varying distances beyond the dust sublimation radius.
They found that the cloud resides in the distance range $r \simeq 0.1-1 \pc$ from the explosion.
At a distance of $0.1 \pc$ they found the width of the cloud (or partial shell) to be $\Delta r \sim 1.4\times 10^{14} \cm$.
Most of the Ca is locked-up in dust grains.

We see the following problems with their model.
\newline
(1) {\it Width of shell/clump.} The inferred expansion velocity of the absorbing material relative to the SN progenitor is $\sim 10 \km \s^{-1}$ \citep{Simonetal2009}.
The expanding gas has a sound speed of $C_{\rm sound} \sim 1(T/100 \K)^{1/2} \km \s^{-1}$.
Even if the mass ejection period lasted for a short time, at $r \sim 0.1 \pc$ the shell width would increase due to its sound speed by about $\sim 3 \times 10^{16} \cm$.
Increasing the gas outflow speed to $100 \km \s^{-1}$ and reducing even more the gas temperature would give a width of $\sim 10^{15} \cm$.
{{{  {We conclude that for any set of parameters compatible with their model and with observations, the radius of such a cloud must be $r > 3 \times 10^{15} \cm$.} }}}
We find a shell width of $\sim 10^{14}$ to be unrealistically small.
For a cloud width of $\sim 3 \times 10^{15} \cm$ and the quoted density the dust optical depth is $\sim 1$, compatible with the extinction toward the SN.
The volume of a clump of radius $\sim 3 \times 10^{15} \cm$ is $V= 10^{47} \cm^{3}$.
If the clump resides at $r \sim 1 \pc$, then its volume is much larger as expansion will bring its size to $ > 10^{16}\cm$ with $V \ga 3 \times 10^{48} \cm^{-3}$.

(2) {\it Mass loss rate in the SD scenario.} The expansion velocity of the cloud relative to the progenitor is $v_{\rm cloud} \simeq 10 \km \s^{-1}$ \citep{Simonetal2009}.
The ratio of the orbital velocity of the AGB star around the center of mass to that of the cloud in the SD scenario is
\begin{eqnarray}
\frac {v_{\rm orb2}}{v_{\rm cloud}} \simeq
1
\left( \frac{v_{\rm cloud}}{10 \km \s^{-1}} \right)^{-1}
\left( \frac{M_{\rm WD}+M_{\rm AGB}}{3.0 M_{\sun}} \right)^{-1/2}
\nonumber
\\
\times
\left( \frac{M_{\rm WD}}{1.4 M_{\sun}} \right)
\left( \frac{a}{5 \AU} \right)^{-1/2},
\label{eq:vorb2}
\end{eqnarray}
where $a$ is their orbital separation.
To eject the cloud into such a small angle $\sim 3 \times 10^{15}\cm/0.1 \pc \sim 0.01$,
the ejection must occur over a time scale of $f \sim 0.01/2 \pi$ of the orbital period. Namely, within about one week.
A cloud of radius $10^{15} \cm$, a width of $1.4 \times 10^{14} \cm $ and the density of their model has a mass of $\sim 10^{-5} M_\odot$.
{{{  {For the radius of the cloud we reduce here for their model, of $\sim 3 \times 10^{15} \cm$, the mass is $\sim 10^{-3} M_\odot$.} }}}
It is not clear how an AGB star can lose such a mass within a week, or a mass loss rate of $> 10^{-3} M_\odot \yr^{-1}$ from a very small surface area.
The WD can launch jets into a very small angle and within a very short time. But they would be much faster than $\sim 10 \km \s^{-1}$.
The conclusion is that any material must occupy a volume of  $V >  10^{47} \cm^{3}$. This is relevant to the next criticism.
Here we note that a larger volume implies a larger mass, which makes the mass loss rate problem more severe.

(3) {\it $H{\alpha}$ emission.} For their hydrogen number density of $n_H=2.4\times 10^7 \cm^{-3}$ and fully ionized gas near maximum, the
$H{\alpha}$ emission should have the following properties.
\newline
(a) Luminosity. The $H{\alpha}$ luminosity near maximum (fully ionized hydrogen) is given by
\begin{equation}
L_{H{\alpha}} \simeq 3 \times 10^{37}
\left( \frac{V}{10^{47} \cm^{3}} \right)
\erg \s^{-1},
\label{eq:macc1}
\end{equation}
where a gas temperature of $\sim 5000 \K$ is assumed \citep{Simonetal2009}.
The minimum volume we derive for the cloud in their model comes for a cloud radius of $\sim 3 \times 10^{15} \cm$ as required by its expansion,
and if it is the only cloud in their model.
{{{  {If the material is in a torus of a large radius $0.1 \pc$ and small radius of $ 3 \times 10^{15} \cm$, the volume is  $5 \times 10^{49} \cm^{-3}$.} }}}
For larger distances from the explosion a lower ionization fraction of hydrogen is possible according to \cite{Simonetal2009}, but then the volume of the clump will
be much larger as we found above.
We conclude that even if the ionization fraction of hydrogen in only $30 \%$ we expected the $H{\alpha}$ luminosity in the model of \cite{Simonetal2009}
to be $L_{H{\alpha}} > 3 \times 10^{37}$
{{{  {as the volume is likely to be more than an order of magnitude larger than the scaling value in equation (\ref{eq:macc1}).} }}}
{{{  {The averaged observed H${\alpha}$ flux is $1.5 \times 10^{16} \erg \cm^{-2} s^{-1}$ from which a luminosity} }}}
of $1.6 \times 10^{37} \erg \s^{-1}$ is derived.
{{{  {The constraint is stronger even, as the H${\alpha}$ has not been changed two years after explosion (J. Simon, private communication 2014),
and hence is not associated with SN 2007le; any H${\alpha}$ from the SN must be $ \ll 10^{37} \erg \s^{-1}$. }  }}}
\newline
(b) As the gas recombines, the $H_\alpha$ luminosity from the clump should drop. The $H_\alpha$ luminosity is not observed to decrease,
and it even slightly  increases by day +84 \citep{Simonetal2009}.
\newline
(c) The $H{\alpha}$ emitting gas should have the same velocity as the Na~I~D lines. However, the $H{\alpha}$ emission is
offset by $16 \km \s^{-1}$ from the velocity of the variable sodium line.
\newline
These three properties show that their model is contradicted by the observed $H{\alpha}$ emission.
{{{  {One way out from this contradiction is if the Na/H ratio is much larger than that in the Sun \citep{Phillipsetal2013}.} }}}

For the problems listed above we find the model proposed by \cite{Simonetal2009} unattractive, and suggest the NaDA model for the variable Na~I~D lines in
SN~2007le.
\cite{YakshinskiyMadey1999} show with laboratory experiments that PSD with wavelength of $\lambda < 3000 \AA$ can account for the lunar release of Na,
with a rate on the surface facing the Sun of  $\dot \phi_{\rm Na} \sim 10^7 {~\rm atoms}~\cm^{-2} \s^{-1}$.
{{{{  {Some less efficient PSD of Na can take place at somewhat longer wavelengths \citep{YakshinskiyMadey2004}.} }}}}
\cite{Leblancetal2008} list PSD as a possible process for the release of Na in comet McNaught C/2006 P1.
If the Na in comets does come from PSD, then the release of Na from dust is more efficient than that from the moon surface, with a flux of
$\dot \phi_{\rm Na} \sim 10^8 {~\rm atoms}~\cm^{-2} \s^{-1}$ at $1 \AU$ from the Sun.
\cite{McClintocketal2008} estimate the PSD from Mercury, that is expected to be a major source of Na to the exosphere,
to be $\sim 2 \times 10^7 \cm^{-2} \s^{-1}$ (on the surface facing the Sun).
At the distance of the moon from the Sun this would be $\sim 4 \times 10^6 \cm^{-2} \s^{-1}$.
The fraction of Na to Ca atoms is $>10$ and the main calcium source on Mercury is not PSD,
but rather micro-meteorite impact vaporization and solar wind induced ion sputtering \citep{Wurzetal2010}, both of which are not relevant to the NaDA model.
In the PSD process assumed in the NaDA model the release of atomic calcium from dust is very low Ca/Na~$ \ll 0.01$.

Near peak luminosity the effective temperature of the SN is almost twice that of the Sun (e.g. \citealt{MaedaIwamoto2009}), implying high UV radiation.
For an effective SN temperature of $10,000 \K$, and taking line blanketing that reduces UV flux by a factor of about 3 \citep{Pauldrachetal1996} {{{{  {or larger,} }}}}
the number of photons with $\lambda < 3000 \AA$ per unit energy is {{{{  {several} }}}} times that of the Sun.
{{{{  {From the spectrum of SN~1992A as given by \cite{Kirshneretal1993} at +5 days past maximum,
we find the photon rate for this UV range to be $\sim 4\times 10^9~{\rm phots} \erg^{-1}$.
This is about equal to the value of UV photon per unit energy from the Sun.
Using the sodium flux from comets, $\dot \phi_{\rm Na} \sim 10^8 {~\rm atoms}~\cm^{-2} \s^{-1}$, quoted above, and considering the large uncertainties, }   }}}}
we scale the rate of PSD of Na from dust at $1 \AU$ form a star {{{{  {with a spectrum as} }}}} that of the SN near peak luminosity with
$\dot \phi_{\rm Na-SN} \simeq 10^8 (L/1 L_{\sun})(r_d/1 \AU)^{-2} {~\rm atoms}~\cm^{-2} \s^{-1}$.

Taking the dust to cover an effective area fraction of $\eta \sim 1$ at radius $r_d$, after a time $t$ the column density of Na~I is
\begin{eqnarray}
N_{\rm Na} \simeq 10^{13} \eta
\left( \frac{r_d}{1 \pc} \right)^{-2}
\left( \frac{L_{\rm SN}}{5\times 10^9 L_\odot} \right)
\nonumber
\\
\times
\left( \frac{\dot \phi_{\rm Na-SN}}{10^8 {~\rm atoms}~\cm^{-2} \s^{-1}} \right)
\left( \frac{t}{10 \days} \right)
 {~\rm atoms}~\cm^{-2} .
\label{eq:nna}
\end{eqnarray}
This falls in the bulk column density range mentioned by \cite{Simonetal2009} for SN~2007le.
{{{{  {For the parameters used above the total sodium mass in a spherical shell is $2.3 \times 10^{-5}(N_{\rm Na}/ 10^{13} \cm^{-2}) M_\odot$.
For a solar composition this corresponds to a total shell mass of $0.7 M_\odot$.
For example, $20 \%$ of the sodium in a shell mass of $3.5 M_\odot$ and solar composition can supply the absorbing gas for the parameters used in equation
\ref{eq:nna} under the assumption of a spherical shell. The mass and/or the fraction of absorbing Na can be lower if the CSM is not in a spherical distribution, as is most likely the case (see below).  } }}}}

We can estimate the shell mass from the properties of the dust. We take grains of size $a_d=0.01 \mu m$ and density of $2 \g \cm^{-3}$ to cover a fraction $\eta$
of the surface area as seen from the SN progenitor. The gas to dust mass ratio is taken to be 100. A spherical shell mass is
\begin{equation}
M_{\rm shell} \sim
20 \eta
\left( \frac{a_d}{0.01 \mu m} \right)
\left( \frac{r_d}{1 \pc} \right)^{2}
M_{\sun}.
\label{eq:mshell}
\end{equation}
{{{{  {The grain size in planetary nebulae can be even smaller, $a_d < 0.01 \mu m$ \citep{Lenzunietal1989}, reducing the required mass. } }}}}
In the NaDA model the source of this mass is either the ISM in the SN vicinity, hence not a complete shell but rather a cloud,
or a planetary nebula shell that was ejected $\sim 10^5$ years before explosion \citep{TsebrenkoSoker2013, TsebrenkoSoker2014}.
In the PN case it is more likely to be a ring with a much smaller volume. A more typical mass is $\la 10 M_\odot$, including a PN shell and the ISM it entrained.
The SN radiation ionizes a very small fraction of the hydrogen in the shell, and the H$\alpha$ luminosity problem  mentioned in section \ref{sec:SN2007le} does not exist.

{{{{  {While sodium is not depleted much to dust in the ISM, it can be depleted to dust in evolved AGB and post-AGB stars, e.g., IRAS~17038-4815 \citep{Maasetal2005}
and IRC+10216 \citep{MauronHuggins2010}. In the later object $20\%$ of Na in the gas phase and $80 \%$ in dust \citep{MauronHuggins2010}.
To protect this dust from destruction in the planetary nebula phase preceding the SN I explosion, the dust must be dense,
implying concentrated in the equatorial plane. We therefore take the dust in the NaDA model to be concentrated in the equatorial plane.
This is compatible with SN Ia showing variable Na~I lines being very rare. As well, we take the sodium to reside {\it both in the dust and gas phases.}  } }}}}


\section{SUMMARY}
\label{sec:Summary}

Examining the previously suggested models to explain time-variable Na~I~D absorbtion lines from SN~2006X (section \ref{sec:SN2006X})
and SN~2007le (section \ref{sec:SN2007le}), we showed that these models suffer from severe problems.
These models assume that variable absorption depth result from variable amount of atomic sodium due to ionization and recombination.
To account for such variations the electron density should be large, which in turn implies large hydrogen density.
Such shells cannot be within the dust sublimation radius \citep{Chugai2008, Simonetal2009}, and hence must be at a distance of
$\ga 0.1 \pc$ from the explosion. The large volume implies a large shell (or ring) mass that cannot be accounted for in the single-degenerate (SD) scenario
for type Ia supernovae (SN Ia).
In the case of SN~2007le such a cloud is expected to have an H$\alpha$ luminosity which is much above the observed luminosity (section \ref{sec:SN2007le}).

We suggest instead that the sodium responsible for the absorption is released from dust grains in the same way as sodium is released from dust grains of comets
that approach the sun to within $\sim 1 \AU$ (e.g., \citealt{Cremoneseetal1997, Leblancetal2008}).
The increase in absorption strength is attributed to more release of Na with time.
The positive correlation between large extinction and variable Na~I~D absorption line strength \citep{Blondinetal2009} can be accounted for
in this Na-from-dust absorption (NaDA) model.

{{{{  {The decrease in absorption, as in the low velocity segment of SN~2006X \citep{Patatetal2007}, is attributed to ionization of neutral sodium,
as in the modelling of \cite{Borkowskietal2009}.
The overall variability of the Na~I~D absorption lines is determined by the competition between the PSD of Na from dust that lead to an increase, and ionization that
causes decrease, in absorption strength. Some of the Na might be in the gas phase to begin with. } }}}}

In the NaDA model the absorbing shell resides at $\sim 0.1-3 \pc$ from the SN,
is most likely concentrated in an equatorial plane, and its mass is $\sim 1-10 M_{\sun}$ (eq. \ref{eq:mshell}).
This is larger than what a regular AGB wind can supply, but is compatible with a planetary nebula shell or a planetary nebula shell that entrained some
interstellar medium (ISM).
The typical mass to be used in the proposed model is $\sim 5 M_{\sun}$.
The idea of a SN Ia within a planetary nebula shell was proposed for the Kepler SN by \cite{TsebrenkoSoker2013},
and for G1.9+0.3 and RCW86 (SN~185) by \cite{TsebrenkoSoker2014}.
Here we speculate that some (or all), SN Ia with variable Na~I~D absorption lines exploded within a planetary nebula shell.

The results here add to the accumulating evidence that whenever a circumstellar medium (CSM) is found around a SN Ia, it cannot be accounted for
by the SD scenario, e.g., as for the SN Ia SN PTF~11kx \citep{Sokeretal2013}.
Contrary to the trend of using any discovered CSM to support the SD scenario for SN Ia, we argue here that these findings actually
contradict the SD scenario.

{{{  {I thank Laura Chomiuk, Nikolai Chugai, Josh Simon, Assaf Sternberg, and two anonymous referees for helpful comments.} }}}

{}

\end{document}